%% file: COPENHAGEN-PROC.TEX
\documentclass[12pt]{article}
\usepackage{epsfig}

\input econfmacros.tex
\def\Title#1{\begin{center} {\Large {\bf #1} } \end{center}}

\begin{document}

\Title{Effective Theory for QCD in the LOFF Phase}

\bigskip\bigskip


\begin{raggedright}

{\it Giuseppe Nardulli\index{Nardulli, G.}\\
Dipartimento Interateneo di Fisica,
Universit\`a di Bari,\\
I.N.F.N., Sezione di Bari\\ I-70126 Bari, ITALY}
\bigskip\bigskip
\end{raggedright}
\newcommand{\be}{\begin{equation}}
\newcommand{\dd}{\displaystyle}
\newcommand{\eequa}{\end{equation}}
\newcommand{\bea}{\begin{eqnarray}}
\newcommand{\eea}{\end{eqnarray}}

\section{Introduction}

Recently, the old idea of color superconductivity \cite{others}
has been given a new life \cite{alford}, after the discovery that
QCD at high density and low temperature can undergo a phase
transition to a color super-conducting state characterized, for
three light quarks, by Color Flavor Locking: CFL (for recent
reviews see \cite{rassegne}). In this talk I wish to examine a
different phase of QCD also present at low temperature and high
baryonic chemical potential \cite{LOFF}; for two flavors this
phase is characterized by isospin breaking and the emergence of a
crystalline pattern of the diquark condensate. In a recent paper
\cite{gattoloff} it has been shown that in this phase there exists
a quasi particle (phonon) associated to spontaneous breaking of
rotational and translational invariance. Its properties can be
studied by an effective lagrangian approach \cite{hong},
\cite{cflgatto}, \cite{casalbuoni}; they will be reviewed here
together with a short discussion on possible astrophysical
implications.

\section{Crystalline colour superconductive phase}In any possible
application of the color superconductivity, for instance in the
inner core of neutron stars, flavor symmetry is likely to be
broken not only explicitly by quark mass terms, but also by weak
interactions. For example in compact stars, considering only two
flavors, isospin is broken by $\delta\mu=\mu_u- \mu_d\neq 0$, due
to the process:
 \be d\to u e \nu\ .
 \eequa
When the two quarks in the Cooper pair have different chemical
potentials, the vacuum is characterized, for certain values of
$\delta\mu$, by a non vanishing expectation value of a quark
bilinear breaking translational and rotational invariance. The
appearance of this condensate is a consequence of the fact that
 in a given range of
$\delta\mu$ \cite{LOFF}, the formation of a Cooper pair with a
total momentum\be \vec p_1\,+\,\vec p_2\,=\,2\vec q\neq \vec
0\eequa
 is energetically favored. A similar phenomenon was already observed many years ago in the
context of the BCS theory for  superconducting materials in
presence of magnetic impurities by Larkin, Ovchinnikov, Fulde and
Ferrel and the corresponding phase is named LOFF state
\cite{originalloff}.

 The exact
form of the order parameter (diquark condensate) breaking
space-time symmetries in the crystalline phase is not yet known.
In \cite{LOFF} the following ansatz is made: \be \Delta(\vec
x)=\Delta e^{i2\vec q\cdot\vec x}=\Delta e^{i2q\vec n\cdot\vec x}\
. \label{ansatz}\eequa The value of $|\vec q|$ is fixed by the
dynamics, while its direction $\vec n$ is spontaneously chosen. In
the sequel I will assume the ansatz (\ref{ansatz}) as well. The
order parameter (\ref{ansatz}) induces a lattice structure given
by parallel planes perpendicular to $\vec n$:\be \vec n \cdot\vec
x\,=\,\frac{\pi k}{q}\hskip 1cm (k\,=0,\, \pm 1,\,\pm 2,...) \
.\label{planes}\eequa We can give the following physical picture
of the lattice structure of the LOFF phase: Due to the interaction
with the medium,  the Majorana masses of the red and green up and
down quarks have a periodic modulation in space, reaching on
subsequent planes maxima and minima. For the Cooper pair to be
formed one needs a color attractive  channel, which is the
antisymmetric channel; therefore it must be in antisymmetric
flavor state if it is in antisymmetric spin state ($S=0$). The
condensate has therefore the form \be
-<0|\epsilon_{ij}\epsilon_{\alpha\beta 3 } \psi^{i\alpha}( \vec
x)C\psi^{j\beta}(\vec x)|0>= 2\Gamma_A^L e^{2i\vec q\cdot\vec
x}\label{scalar}\ .\eequa The analysis of \cite{LOFF} shows that,
besides the condensate (\ref{scalar}) (scalar condensate), another
different condensate is possible, i.e. one characterized by total
spin 1 (vector condensate) and by a symmetric flavor state: \be
i<0|\sigma^1_{ij}\epsilon_{\alpha\beta 3 } \psi^{i\alpha}(\vec
r)C\sigma^{03}\psi^{j\beta}(\vec x)|0>=2\Gamma_B^L e^{2i\vec
q\cdot\vec x}\ .\label{vector}\eequa Note that in the BCS state
the quarks forming the Cooper pair have necessarily $S=0$.

It goes without saying that the hypotheses of  \cite{LOFF} are
rather restrictive, as these authors assume only two flavors and
make the ansatz of a plane wave behavior, Eq. (\ref{ansatz}). In
any event their results are as follows. Assuming a pointlike
interaction as the origin of the Fermi surface instability, the
LOFF state is energetically favored in a small range of values of
$ \delta\mu$ around $\delta\mu\sim 0.7\Delta$. The actual value of
the window range compatible with the presence of the LOFF state
depends on the calculation by which the crystalline color state is
computed. While the small interval is based on a local
interaction, assuming gluon exchange, as in \cite{LOFFbis}, the
window opens up considerably.

The order parameters (\ref{scalar}) and (\ref{vector})
spontaneously break rotational and translational symmetries.
Associated with this breaking there will be Nambu Goldstone Bosons
(NGB) as in a crystal; these quasi-particles are known as phonons.
Let us discuss them in some detail.

\section{Phonons in the LOFF phase}
\begin{figure}[htb]
\begin{center}
\epsfig{file=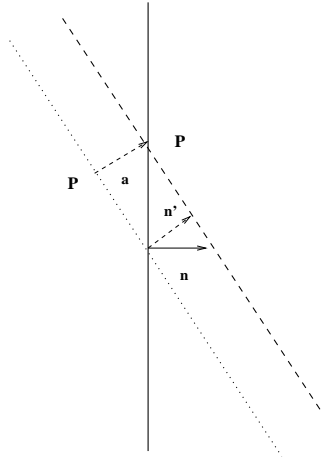,height=2.4in} \caption{In the point P the
effect of the rotation $\vec n\to\vec n^{\,\prime}$ and the effect
of the translation $\vec x\to \vec x+\vec a$ tend to compensate
each other.} \label{fig:phonon3}
\end{center}
\end{figure}

For a generic lattice structure it is known that there are three
phonons associated to the breaking of space symmetries. However
one can show \cite{gattoloff} that in order to describe the
spontaneous breaking of space symmetries induced by the
condensates (\ref{scalar}) and (\ref{vector}) one NGB is
sufficient. The argument (see Fig. \ref{fig:phonon3}) is as
follows: Rotations and translations are not independent
transformations, because the result of a translation plus a
rotation, at least locally, can be made equivalent to a pure
translation.
The NGB is a long wavelength small amplitude variation of the
condensate $\Delta(\vec x)\to \Delta(\vec x)e^{i\phi/f}$, with
 \be
 {\phi}/{f}\,=\,2q(\vec n+\delta\vec n)\cdot(\vec x+
 \delta\vec x)-2q\vec n\cdot\vec x\,=\,2q\vec R\cdot\vec x+ T-2q\vec n\cdot\vec x\ ,\label{35}\eequa
 where we have introduced  the auxiliary function $T$,
  given by $ T\,=\,2\,q\,\vec R\cdot\delta\vec x$.

Now the lattice fluctuations $\phi/{f}$ must be small; from this
it
  follows that $T$ and $\vec R$ are not independent fields and therefore $T$ must depend functionally on $\vec R$, i.e. $T=F[\vec R]$,
 which, using again (\ref{35}), means
  that\be \Phi\equiv2q\vec n\cdot\vec x+{\phi}/{f}
 =2q\vec R\cdot\vec x+{F[\vec R]}\equiv G[\vec R,\,\vec x]\ .\eequa
 The solution of this functional relation has the form
 \be \vec R=\vec
 h[\Phi]\eequa where $\vec h$ is a vector built out of the scalar function
 $\Phi$. By this function one can only\footnote{In principle
there is a second vector,  $\vec x$, on which $\vec R$ could
depend linearly, but this possibility is excluded because $\vec R$
is a vector field transforming under translations as $\vec R(\vec
x)\to \vec R^{\,\prime}(\vec x^{\,\prime})=\vec R(\vec x)$.} form
the
 vector $\vec\nabla \Phi$ ; therefore  we
 get
 \be
\vec R =\frac{\vec \nabla\Phi}{|\vec \nabla\Phi|}\,
,\label{r3}\eequa which satisfies $|\vec R|=1$ and $<\vec
R>_0=\vec n$. In terms of the phonon field $\phi$ the  field $\vec
R$ is given at the first order in $\phi$ by the expression\be \vec
R=\vec n+\frac{1}{2fq}\left[\vec\nabla\phi-\vec n(\vec n\cdot\vec
\nabla\phi)\right]\,.\label{63}\eequa The interaction term between
the NGB  and the quarks   is contained in \be {\cal
L}_{int}=-\frac 1 2\, e^{i\Phi} \sum_{\vec
v}\left[\Delta^{(s)}\epsilon_{ij}+\Delta^{(v)}(\vec v\cdot\vec R
)\sigma^1_{ij} \right]\epsilon^{\alpha\beta 3}\psi_{i,\alpha,\vec
v}\,C\,\psi_{j,\beta,-\vec v}+h.c.-(L\to R)\ ;
\label{lextrnal}\eequa from this lagrangian, following the same
bosonization procedure of \cite{cflgatto}, one can obtain the
effective lagrangian for the massless scalar  field $\phi$
\cite{gattoloff} and derive its dispersion law. It is worth
observing that the fields $\psi_{j,\beta,\pm\vec v}$ in
(\ref{lextrnal}) are velocity dependent fields; they are computed
at opposite velocities, as off-diagonal terms in the average over
velocities cancel out.
\smallskip
\section{Glitches and the LOFF phase}
The pulsars are rapidly rotating stellar objects, characterized by
the presence of strong magnetic fieds and by an almost continuous
conversion of rotational energy into electromagnetic radiation.
The rotaton  periods can vary in the range $10^{-3}$  sec up to a
few seconds; these periods increase slowly and  never decrease
except for occasional  glitches, when the variation of frequency
can be   $ \delta\Omega/\Omega\approx 10^{-6}$ or smaller.

\begin{figure}[htb]
\begin{center}
\epsfig{file=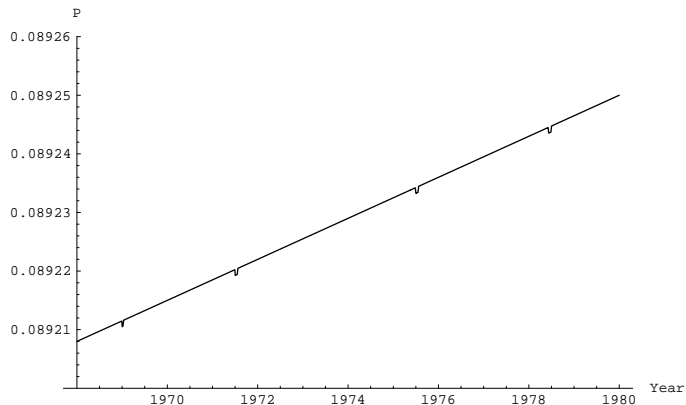,height=2.2in} \caption{Schematic view
of the period $P$ of the PSR 0833-45 (Vela) showing four glitches
in the years 1969-1980.} \label{fig:vela}
\end{center}
\end{figure}

Glitches are a typical phenomenon of the pulsars, in the sense
that probably all the pulsar have glitches. The ordinary
explanation of the glitches is based on the idea that these sudden
jumps of the rotational frequency  are due to the movements
outwards of rotational vortices in the neutron superfluid and
their interaction with the crust. This is one of the main reasons
that allow the identification of pulsars with neutron stars, as
only neutron stars are supposed to have a metallic crust.
 A schematic  diagram
of glitches is shown in Fig. \ref{fig:vela}

The interesting aspect of the LOFF phase discussed in this talk is
that even in quark stars, provided one is in a color
superconductivity phase, one can have a crystal structure given by
a lattice characterized by a geometric array where the gap
parameter varies periodically. Therefore also these stars might
have glitches and the  possibility arises that some of the pulsars
are indeed quark stars \cite{bombaci}. In a more conservative vein
one can also imagine that the LOFF phase be realized in the inner
core of neutron star; in this case the crystalline color
superconductivity could be partly responsible for the glitches of
the pulsar. A detailed analysis  of this scenario is however
premature as one should first complete the study of the LOFF phase
in two directions, first by including the strange quark  and,
second, by sorting out the exact form of the color lattice.

\bigskip
I am grateful to R. Casalbuoni, R. Gatto, M. Mannarelli for their
precious collaboration in the work on which this contribution is
based and to I. Bombaci and  K. Rajagopal for enlightening
discussions on the subject of my talk.

\def\Discussion{
\setlength{\parskip}{0.3cm}\setlength{\parindent}{0.0cm}
     \bigskip\bigskip      {\Large {\bf Discussion}} \bigskip}
\def\speaker#1{{\bf #1:}\ }
\def\endDiscussion{}

\Discussion

\speaker{J. Madsen (University of Aarhus)} How would you visualize
the LOFF phase? I.e.: What are the "ions" in the crystal and how
are they organized?

\speaker{Nardulli}  In the LOFF phase the condensate varies
periodically in space with minima and maxima. Maxima are places
where the condensate is bigger and therefore the fermions have the
largest Majorana mass. The lattice of "ions" could  be visualized
as parallel planes where the Cooper pairs are more tightly
bounded.

\speaker{D. K. Hong (Pusan National University)}  When you derived
the effective theory of the LOFF phase, you used the same Fermi
velocity for up and down quarks. But it would be much easier to
analyze the effective theory using different Fermi velocities that
satisfy the constraint $\mu_u\vec v_u+\mu_d\vec v_d=2\vec q $.

\speaker{Nardulli}  The use of one velocity has the advantage of a
Fermi velocity superselection rule, which reduces the complexity
of the loop integrals. It arises because in the $\mu\to\infty$
limit the rapid oscillations of the exponential factor in the
lagrangian cancel all the contributions except those satisfying $
\vec v_u+\vec v_d=0$.
\endDiscussion
\end{document}

%% file: econfmacros.tex
\def\beq{\begin{equation}}
\def\eeq#1{\label{#1}\end{equation}}
\def\eeqn{\end{equation}}

\def\beqa{\begin{eqnarray}}
\def\eeqa#1{\label{#1}\end{eqnarray}}
\def\eeqan{\end{eqnarray}}

\let\bar=\overbar

\def\Dslash{\not{\hbox{\kern-4pt $D$}}}
\def\dslash{\not{\hbox{\kern-2pt $\del$}}}

\def\msb{{\bar{\ssstyle M \kern -1pt S}}}

